\documentclass[defaultstyle,12pt]{article}

\usepackage{latexsym} 
\usepackage{graphicx}       
\usepackage{rotating}      
\usepackage{amssymb}
\usepackage{amsmath} 

\begin{document}

\title{Internal nonlinear transmission in an Yb mode-locked fiber laser through bifurcations}

\author{Cec\'{i}lia L. A. V. Campos, Lucas B. A. M\'{e}lo,\\
Hugo L. D. de S. Cavalcante, L\'{u}cio H. Acioli, and\\
Marcio H. G. de Miranda\\
\textit{Departamento de F\'{i}sica, Universidade Federal de Pernambuco}\\
\textit{50740-560 Recife, PE - Brazil}\\
\textit{Departamento de Inform\'{a}tica, Centro de Inform\'{a}tica,}\\
\textit{Universidade Federal da Para\'{i}ba}\\
\textit{58051-900 Jo\~{a}o Pessoa, PB - Brazil}\\
\textit{mhgm@df.ufpe.br}\\
Published in Optics Communications, \textbf{26} 10, 13686 (2019).}

\maketitle

\begin{abstract}
Mode-locked fiber lasers are rich dynamical systems that may present several different types of pulsed operating modes, depending on external control parameters such as pump power. A systematic experimental characterization of such regimes is a challenging problem. Here single pulse regimes of ytterbium mode-locked fiber laser are explored and related to the nonlinear amplitude modulation affecting the pulses inside the laser cavity. To understand the full dynamics of this system, average power measurements (slow time scale) have been performed and compared to simultaneous time-resolved measurements (fast time scale). The average nonlinear transmission resulting from the amplitude modulation is measured, allowing to relate the transitions between dynamical regimes to the system's nonlinearity.
\end{abstract}

\section{Introduction}

The interest in ytterbium-doped mode-locked fiber lasers (MLFL) has increased over the last decades because, beyond been a relatively cheap system, it has important technological applications in the fields of physics, engineering, biology and medicine. It is also very interesting from the dynamical systems  point of view because it presents a dense forest of different phenomena to be explored such as: deterministic chaos\ \cite{melo2018deterministic}, rogue waves~\cite{runge2014raman, liu2015rogue}, soliton explosion~\cite{runge2015observation}, dissipative soliton molecules~\cite{grelu2012dissipative}, harmonic mode-locking~\cite{Zhou:06} and multi-pulsing regime~\cite{li2010geometrical, komarov2005multistability, haboucha2008mechanism}.

The Yb MLFL is quite flexible in the sense that the system operates in the pulsed regime under very different conditions. 
One is allowed to externally control two basic properties which determine the operating regime: (\textit{i}) group velocity 
dispersion (GVD) $\beta_2$,  and (\textit{ii}) the cavity nonlinear amplitude modulation (nonlinear loss or saturable absorber). The parameter $\left.\beta_2\right|_{\text{fiber}}$ of most optical fibers is positive at $\lambda\approx1~\mu$m, which is the Yb lasers emission. For Yb laser emission most optical fibers present parameter $\left.\beta_2\right|_{\text{fiber}}>0$. Even so, using a grating pair, one can set the overall dispersion in the cavity at will. To obtain a stable mode-locked pulse train, the GVD is usually set to $\beta_2 \lesssim 0$, but short pulse generation from Yb fiber lasers with overall normal dispersion, $\beta_2 >0 $ regime has been demonstrated in references~\cite{chong2007all, ortacc2009approaching}.

The presence of nonlinear amplitude modulation inside the laser cavity is a fundamental requirement to obtain mode-locked operation~\cite{Nelson1997, Haus:91}, and in many MLFL the most common choice to obtain intensity-dependent losses is based on the nonlinear polarization rotation (NPR)~\cite{Hofer:91}, due to the nonresonant $\chi^{(3)}$ of the optical fiber. This emulates a fast saturable absorber, which has been extensively studied in the operation of pulsed femtosecond lasers. In fiber lasers this is achieved by a combination of waveplates and polarizers to obtain a nonlinear loss mechanism of the form
\begin{equation}
L(P) = L_0 -\gamma\,P\,,
\label{eq:nonlinear_loss}
\end{equation}
\noindent
where $P$ is the instantaneous power, $L_0$ is the overall linear cavity loss and $\gamma$ is the nonlinear loss coefficient. Suitable settings of waveplates allow one to control both $L_0$ and $\gamma$. The inclusion of higher order terms in Eq.~(\ref{eq:nonlinear_loss}) is essential for obtaining stable mode-locked operation~\cite{Jones:10}, and also known to be important to achieve different dynamical regimes~\cite{Newbury}. 

For a laser operating in the mode-locked regime due to NPR, the losses inside the cavity are directly related to the nonlinear transmission occurring in a polarizing element, such as a polarizing beamsplitter. Notice that the element itself behaves linearly, with the nonlinearity originated in the fiber before it.

The nonlinearities present in MLFL provide a fertile ground for experimental exploration, and in order to understand their origin, one option has been to rely on theoretical models based on the complex Ginzburg-Landau equation (CGLE) or the nonlinear Schr\"{o}dinger equation. This approach has been used to study pulsating and chaotic solitons, period doubling, and pulse coexistence in mode-locked lasers~\cite{akhmediev2001pulsating}. Wei {\it et al.}~\cite{wei2017general} have exploited the nonlinear amplitude modulation to understand the physics behind the stochasticity, hysteresis and multistable regimes. Komarov {\it et al.}~\cite{komarov2005multistability} propose to study a multistable Yb mode-locked laser focusing on the roles of group velocity dispersion and the gain saturation. 

A different approach was proposed by Li {\it et al.}~\cite{li2010geometrical}, who have investigated the importance of the nonlinear transmission to explain the transition between single pulse and multi-pulse regimes within the fundamental repetition period of a mode-locked laser. Our interest here is to experimentally characterize different dynamical regimes, beyond stable mode-locked operation of an Yb MLFL, and their relation to the nonlinear cavity transmission. This is done by measuring average power values (slow time scale) over the whole pulse train and comparing to time-resolved measurements (fast time scale). We have discovered that important details about the nonlinear dynamics happening at the fast time scale can be revealed by looking at the slow time scale parameters. This is interesting and worthy of attention because slow detectors are much more accurate and accessible than fast ones, so we suggest that the nonlinear transmission curve can be an alternative for monitoring stability and dynamical transitions of systems similar to the one presented here. 

In this work, we have measured the nonlinear transmission curve for pulse trains in an Yb MLFL, presenting a clear experimental signature of dynamical regime transitions. To measure the nonlinear transmission curve, we have monitored laser beams inside and outside of the laser cavity, which we will refer as intracavity and extracavity, respectively. We have also measured the average laser power, RF and optical spectra, and the temporal series. The average laser power intracavity and extracavity presented a counter intuitive tendency, where it does not grow linearly with the pump current but has a completely sharp variation in slope due to the transition between dynamical regimes. When both powers are added, they show a complementary behavior, giving a linear growth.

\section{Experimental setup}

\begin{figure}
    \centering
\includegraphics[width=0.7\textwidth]{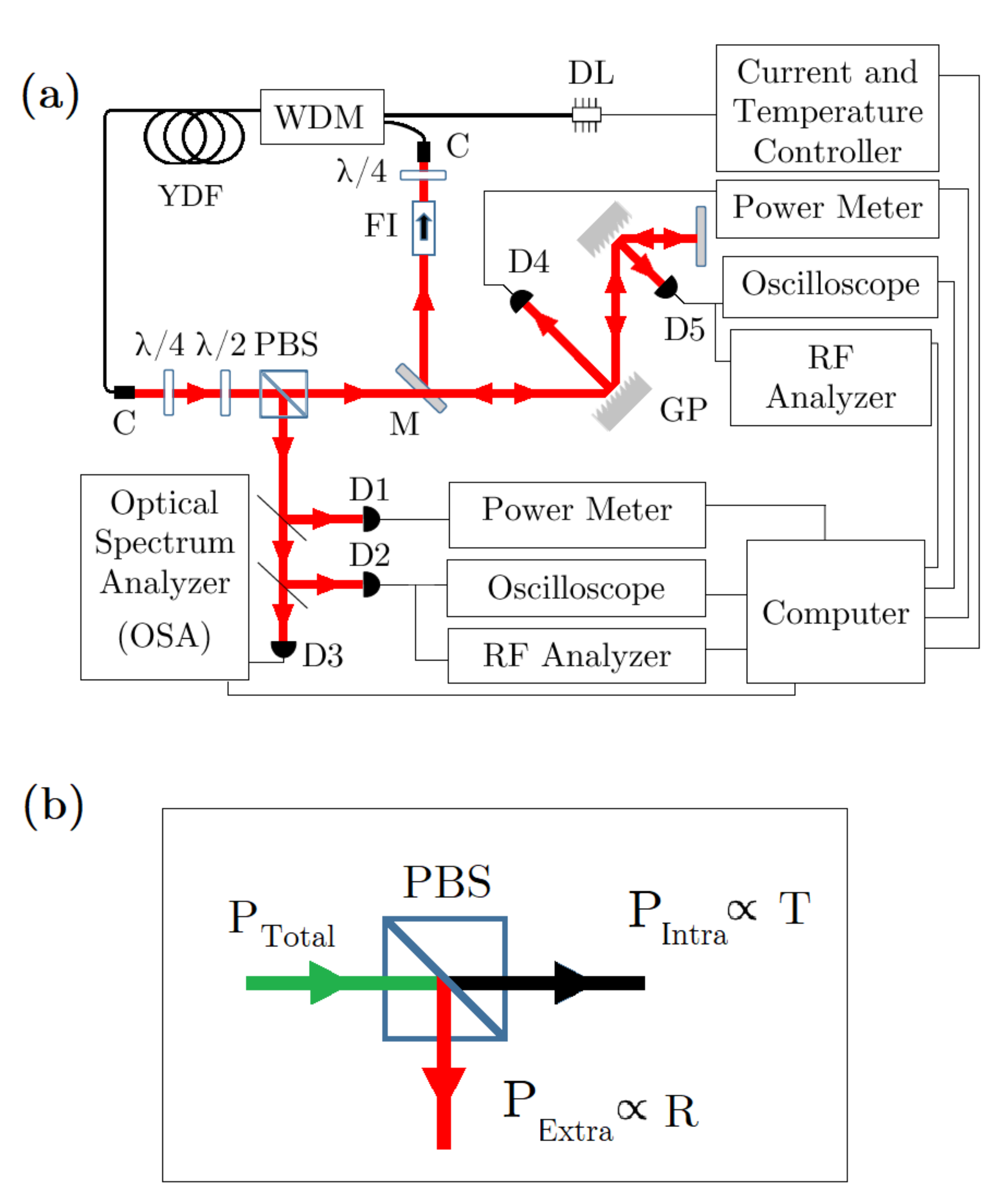}
\caption{(a) Schematic representation of the experimental setup: diode laser (DL), wavelength division multiplexing (WDM), ytterbium doped fiber (YDF), quarter-wave plate ($\lambda$/4), half-wave plate ($\lambda$/2), polarizing beamsplitter (PBS), grating pair (GP), Faraday isolator (FI), GRIN collimators (C), power meter's detectors (D1 and D4), 1 GHz photodetectors (D2 and D5) and pick-up fiber for the OSA (D3). (b) Simple diagram of the laser beams in the PBS.}
\label{f1}
\end{figure}

The experiment was performed using a homemade ytterbium-doped MLFL, similar to the one used in Ref.~\cite{melo2018deterministic}. A schematic diagram of the experimental setup and data acquisition system can be seen in Fig.~\ref{f1}(a). The pump source is a diode laser operating at 976 nm, with maximum power of 720 mW. The gain medium is a highly Yb-doped single-mode fiber CorActive Yb214, with absorption coefficient of 1348 dB/m at 976 nm and length of 22 cm. All the other fibers are standard single-mode HI 1060. The pump power is coupled to the gain medium through a wavelength division multiplexing (WDM), presenting a coupling efficiency of 89$\%$. A pair of GRIN collimators couple the beam to and from the free-space section and provides optical feedback. The coupling efficiency back to the WDM is 35$\%$. The intracavity dispersion is controlled by a grating pair with 600 grooves/mm, each grating presenting a efficiency of 65$\%$ per pass. An optical isolator guarantees the unidirectionality of the signal. The laser is passively mode-locked via NPR technique~\cite{Hofer:91}. This technique is based on the intensity-dependent evolution that a polarization state can exhibit during its propagation in the fibers, provided that its intensity is sufficiently high. The introduction of a polarizer into the cavity will cause an intensity dependent loss, emulating a fast saturable absorber and allowing mode-locking regime to occur. To adjust the polarization and achieve mode-locked operation, a set of waveplates and a polarizing beamsplitter (PBS) are included (Fig.~\ref{f1}(a)). It is important to note that the PBS is also the output coupler of the laser, meaning that the intracavity and the extracavity beams are linearly polarized and orthogonal. The waveplates and PBS losses are very small and will be neglected in the following discussion. When mode-locked, our laser presents a train of pulses with repetition rate of 125 MHz, optical spectrum centered in 1025 nm and temporal width of approximately 200 fs~\cite{melo2018deterministic}.

The experiment consists in scanning the diode pump laser current ($I_{\text{pump}}$) from 525 mA to 969 mA, in steps of 4 mA, keeping the waveplates settings and dispersion fixed. The calibration from $I_{\text{pump}}$ to optical power is 0.68 mW/mA. Data acquisition was automated and controlled by a custom LabVIEW program, involving two photodetectors (1 GHz bandwidth), two RF spectrum analyzers (3 GHz bandwidth), two power meters, one optical spectrum analyzer (OSA) and one oscilloscope (1 GHz bandwidth). 

The time-resolved measurements are temporal series of pulses acquired by fast photodetectors (1 GHz bandwidth). We denote this as a fast time scale measurement. The average power measurements were acquired in a slow time scale, since the power meters have a bandwidth of approximately 50 Hz. For each $I_{\text{pump}}$, intracavity and extracavity time series, RF spectra and average powers were acquired. The optical spectrum was measured after the laser output. All time series were saved directly from the oscilloscope in 200 ns windows and, for each $I_{\text{pump}}$, 20 intracavity and extracavity windows were acquired and processed. It is important to note that the photodetectors are much slower than the femtosecond pulses, but the peak values of the time series are proportional to the pulse energy. It still faithfully represents the periodic nature of the dynamics as shown in our previous work~\cite{melo2018deterministic}.

As stated previously, our main interest is in characterizing the nonlinear transmission through the polarizing beamsplitter. In order to do so, we measure average powers after its two exit ports, as shown in Fig.~\ref{f1}(b). The beam reflected by the PBS is the output of the laser and detector D1 measures its power, as shown in Fig.~\ref{f1}(a). On the other hand, it is not possible measure the power transmitted through the PBS directly, as this would prevent the laser from operating. To circumvent this problem we use the zero order reflection from one of the gratings, collected by detector D4. Careful measurements of the conversion factors for the detectors were performed and we determine that the actual powers immediately after the PBS must be multiplied by 1.3 and 17.9 for D1 and D4, respectively. The internal losses in the PBS are negligible and the total average power that leaves the gain medium is given by the from the sum of reflected and transmitted average powers.

\section{Results and discussion}

\begin{figure}[h]
    \centering
\includegraphics[width=0.70\textwidth]{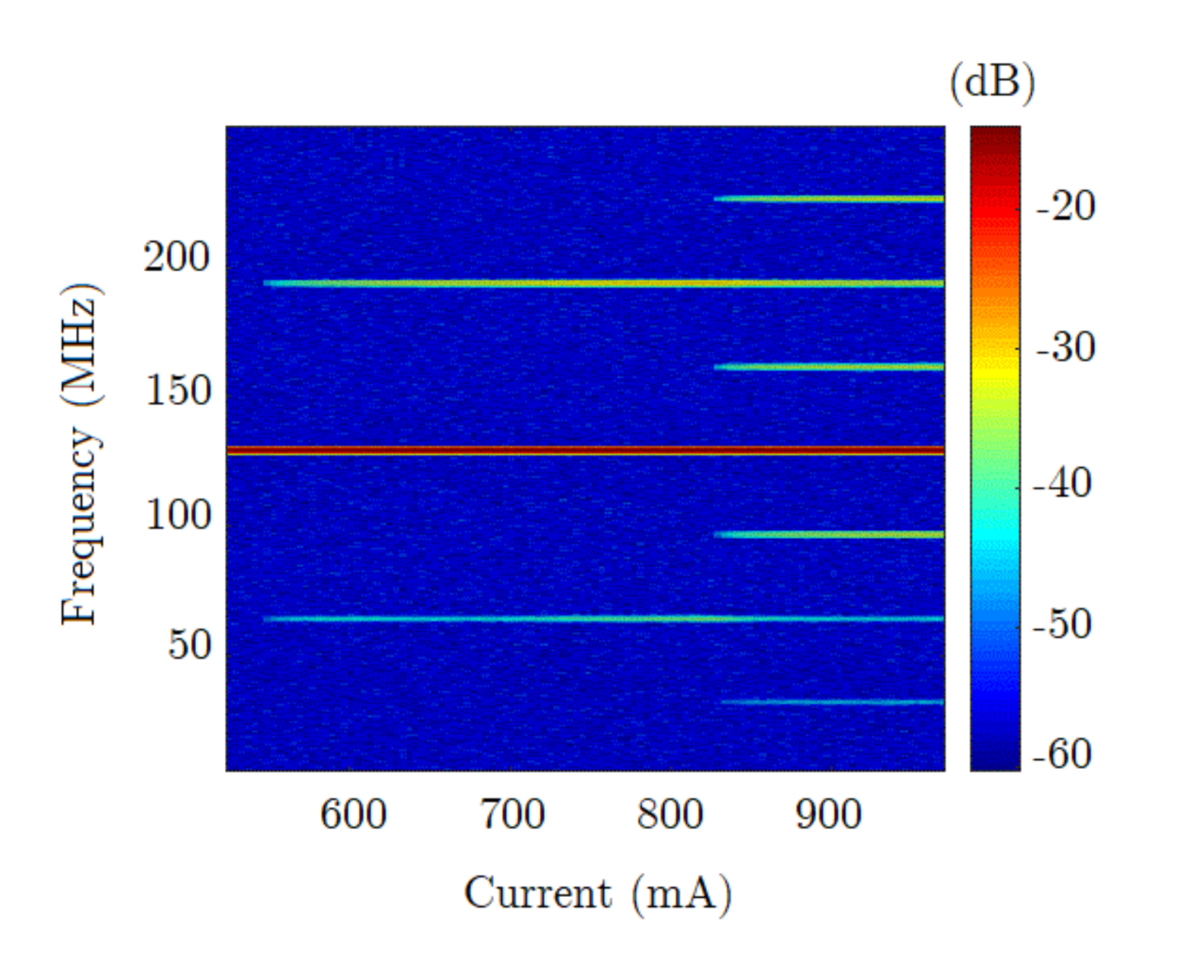}
\caption{Evolution of the extracavity RF spectrum showing period-one, period-two and period-four.}
\label{f2}
\end{figure}

Using $I_{\text{pump}}$ as a control parameter, different dynamical behaviors are observed by looking at the variation of the amplitude of the pulses in a time series. To investigate distinct operating regimes, the evolution of the RF spectrum was systematically registered for intracavity and extracavity signals. As both presented similar dynamics, Fig.~\ref{f2} presents only the extracavity one. For 525 mA $<I_{\text{pump}}<$ 545 mA, there is only one single peak at 125 MHz and it corresponds to a regular mode-locked regime, where pulses have the same amplitude. For 545 mA $<I_{\text{pump}}<$ 829 mA, a new peak arises at approximately 62.5 MHz, half the fundamental frequency, corresponding to period doubling. For 829 mA $<I_{\text{pump}}<$ 969 mA, another peak becomes present at approximately 31.25 MHz, one quarter the fundamental frequency, corresponding to period quadrupling. All the others visible frequencies in the RF spectrum are higher order beat notes of these ones.

\begin{figure*}
\includegraphics[width=1.0\textwidth]{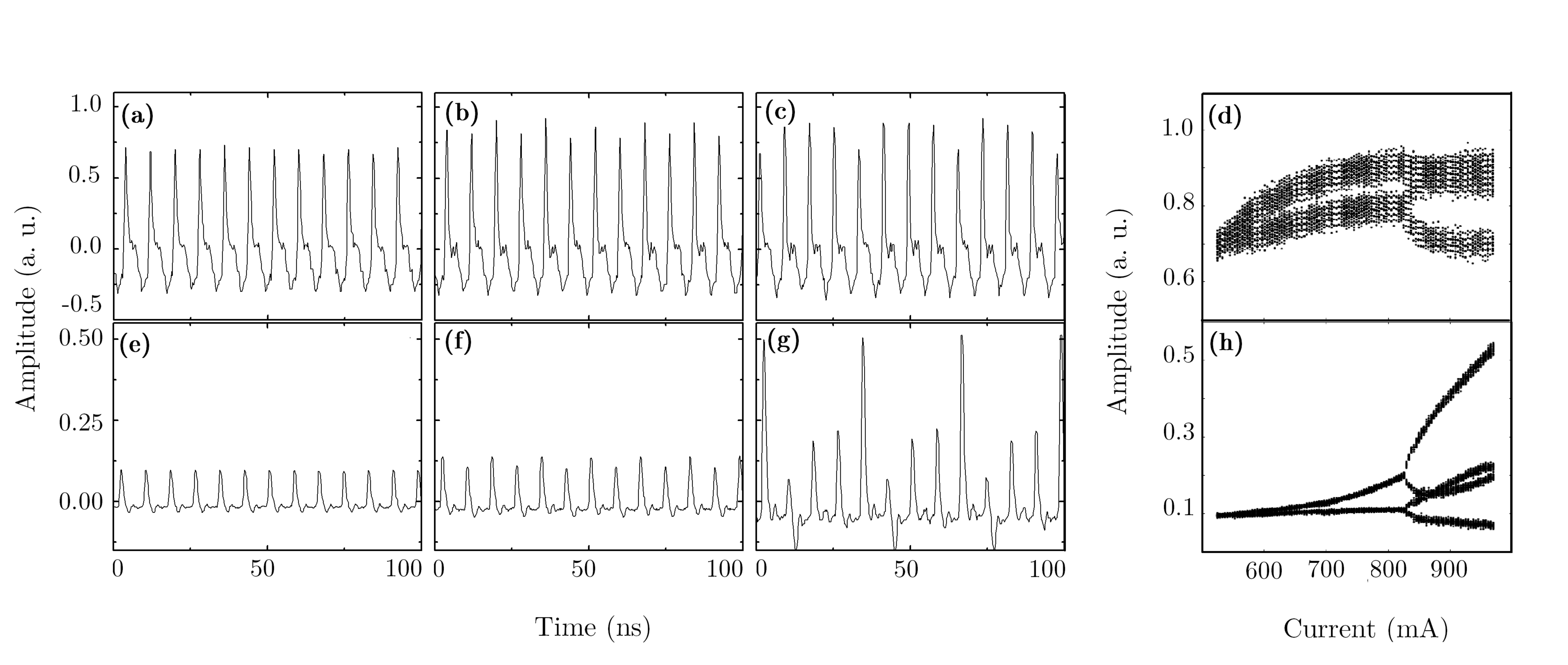}
\caption{Intracavity (upper row) and extracavity (lower row) time series (fast time scale measurements): (a) and (e) period-one (529 mA), (b) and (f) period-two (721 mA) and (c) and (g) period-four (957 mA). Bifurcation diagrams (d) intracavity and (h) extracavity.}
\label{f3}
\end{figure*}

The intracavity and extracavity time series also confirm the nonlinear dynamics observed in the evolution of the RF spectrum. This is seen in the representative time series shown in Fig.~\ref{f3}, where the upper row refers to the intracavity signal and the lower row refers to the extracavity one. The regular mode-locked regime appears in Fig.~\ref{f3}(a) and~\ref{f3}(e), each of them presenting pulses with constant peak values. However, this uniformity does not exist either in Fig.~\ref{f3}(b) or~\ref{f3}(f), where the amplitude of the pulses alternates between two values presenting a period-two regime. Similarly, it happens to the series presented in Fig.~\ref{f3}(c) and~\ref{f3}(g), where the amplitude is modulated among four values, showing a period-four regime. Fig.~\ref{f3}(g) draws attention to the existence of pulses much higher than the others, a very different pattern from the other series. For every four pulses, the largest of them has the amplitude almost ten times bigger than the smallest one, making it more susceptible to the nonlinearity of the gain medium. This pattern occurs throughout the extracavity period-four regime, which exhibits a much higher modulation than the intracavity one shown in Fig.~\ref{f3}(c), being relevant to the analysis because amplitude modulation has a direct relation to the nonlinear transmission curve. All these regimes are better visualized in the bifurcation diagrams presented in Fig.~\ref{f3}(d) and ~\ref{f3}(h), where the peak values from the time series are plotted versus $I_{\text{pump}}$ as control parameter. The branches of the intracavity diagram in Fig.~\ref{f3}(d), remain closer than the branches of the extracavity in Fig.~\ref{f3}(h). Notice that in Fig.~\ref{f3}(d) the pulse amplitude seems to saturate, which implies in higher extracavity amplitudes, as shown in Fig.~\ref{f3}(h). The lack of resolution of the intracavity bifurcation diagram in Fig.~\ref{f3}(d) is due to the smaller difference between the intensities of the time series peaks. 

\begin{figure}[h!]
    \centering
\includegraphics[width=0.70\textwidth]{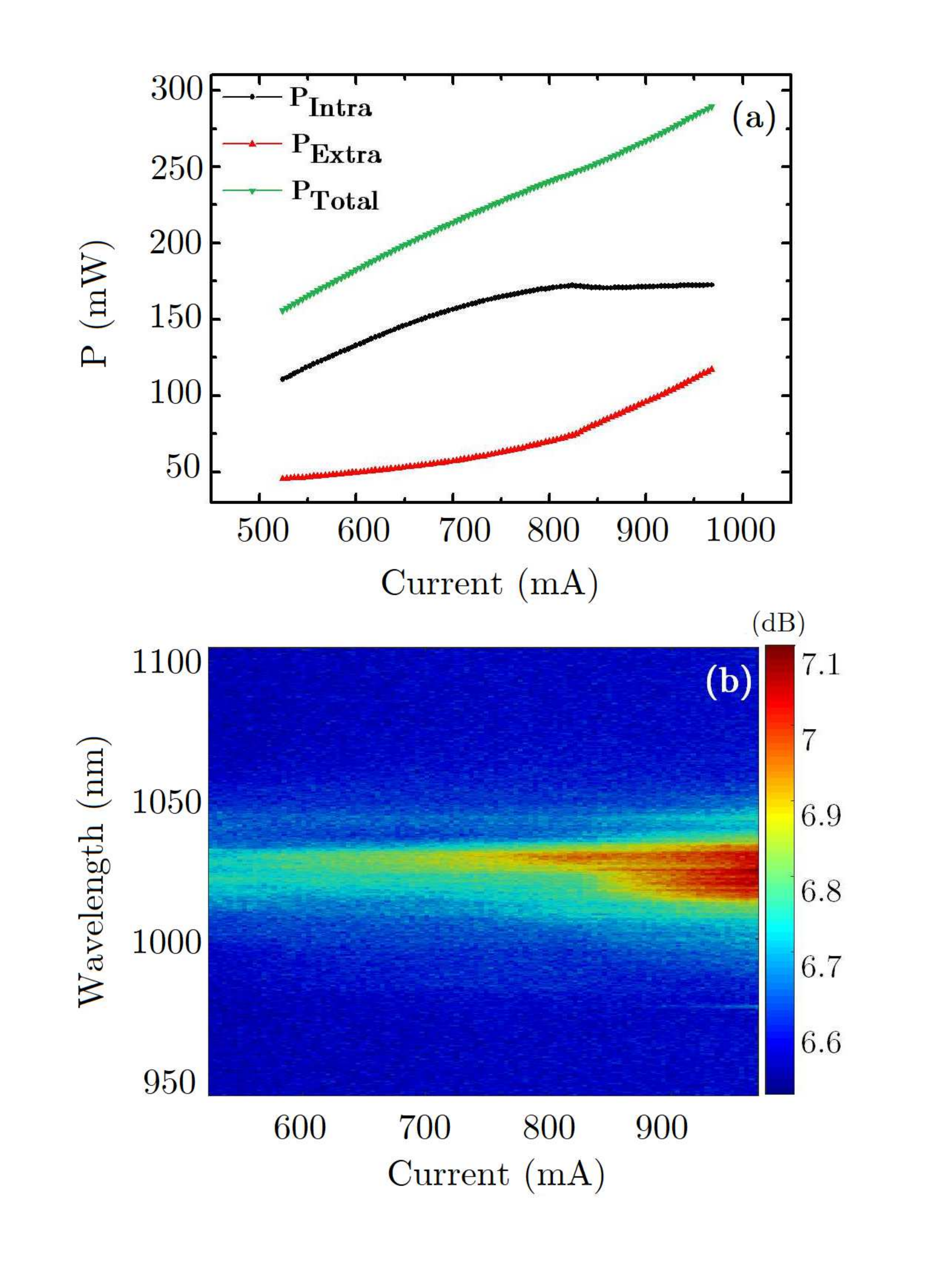}
\caption{(a) Intra, extra and total average power (slow time scale measurements). (b) Evolution of extracavity optical spectrum.}
\label{f5}
\end{figure}

\begin{figure}[h!]
    \centering
\includegraphics[width=0.70\textwidth]{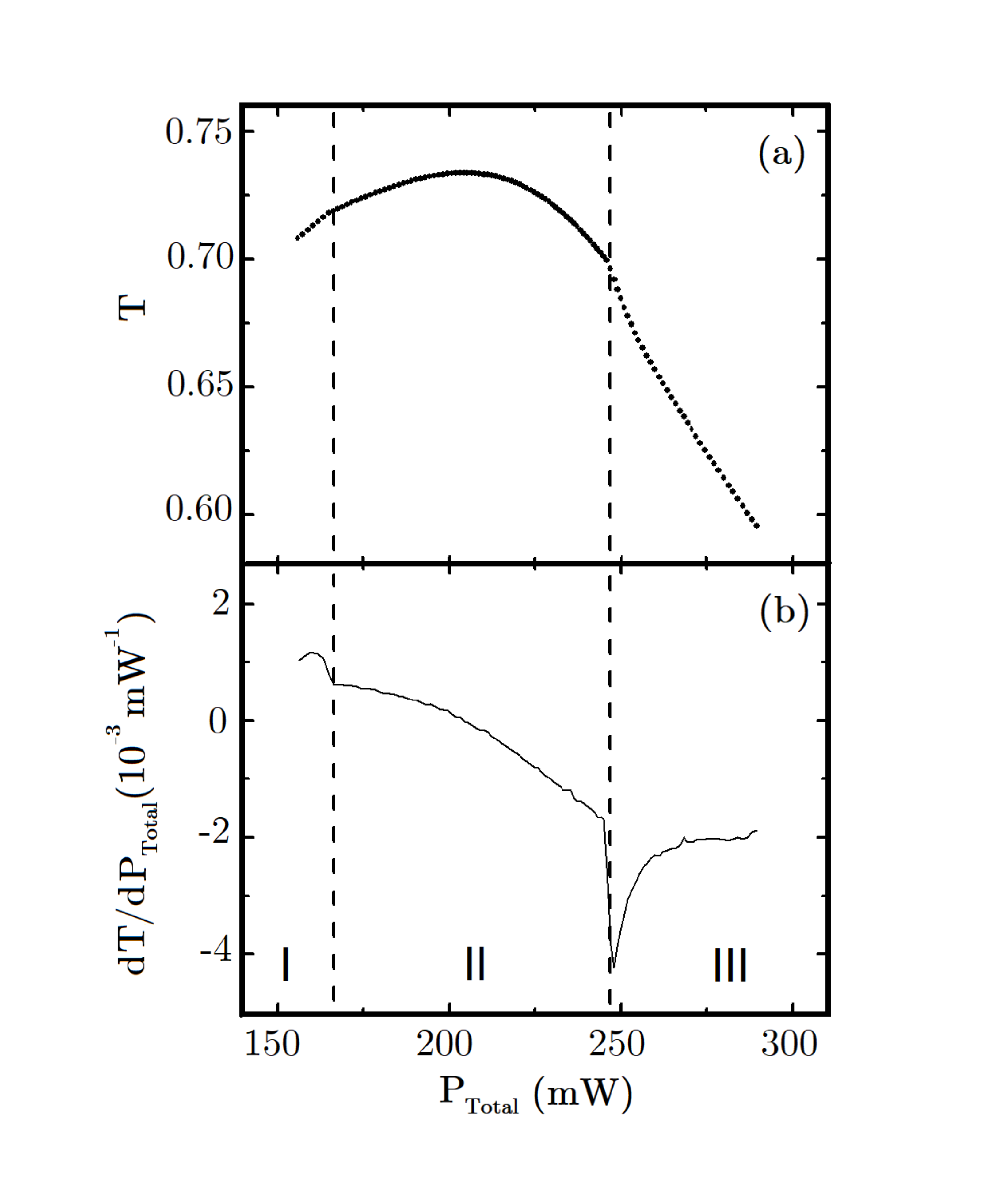}
\caption{Nonlinear transmission curve: (a) PBS transmission $T$ versus $P_{\text{Total}}$, (b) First derivative of $T$ versus $P_{\text{Total}}$.}
\label{f6}
\end{figure}

In order to clarify the meaning of the next set of data, we describe the relation between the total nonlinear cavity loss, $L(P)$ defined in Eq.~(\ref{eq:nonlinear_loss}), and the nonlinear transmission, $T(P)$ at the PBS. Figure~\ref{f1}(b) shows the relevant input and output ports of the PBS. In order to achieve mode-locked operation, the optical feedback must favor higher powers and the portion of light reflected from the PBS must decrease with power. This means the nonlinear reflectivity should be of the form $R(P) = R_{\text{PBS}} -\gamma\,P$. As the internal losses in the PBS are negligible, its nonlinear transmission is $T(P) = 1-R(P)$, which may be rewritten as, 
\begin{equation}
T(P)= 1 - R_{\text{PBS}} + \gamma\,P \equiv T_{\text{PBS}} + \gamma\, P\;,
\label{eq:NLR}
\end{equation}
\noindent
which increases with power, as desired. Importantly, we emphasize that both $R_{\text{PBS}}$ and $\gamma$ may be controlled externally by proper setting of the waveplates. From Eq.~(\ref{eq:nonlinear_loss}), the total intracavity loss in the laser is given by, 
\begin{equation}
L(P) = L_{\text{PBS}} + L_{\text{CAV}} - \gamma\;P,
\label{eq:NLLoss}
\end{equation}
\noindent
where $L_{\text{PBS}} = R_{\text{PBS}}$ and the overall cavity linear loss is $L_0 = L_{\text{PBS}} + L_{\text{CAV}}$, where $L_{\text{CAV}}$ represents all the other linear cavity losses, such as coupling losses back into the fiber, grating losses, etc.

From Fig.~\ref{f1}(b) the nonlinear reflectivity is obtained from measurements of the intracavity and extracavity powers, $P_{\text{Intra}}$ and $P_{\text{Extra}}$, respectively. The total incident power is $P_{\text{Total}} = P_{\text{Intra}} + P_{\text{Extra}}$ and the nonlinear reflectivity and transmission are
\begin{subequations}
\begin{align}
\label{eq:NLTPa}
R(P) = \frac{P_{\text{Extra}}}{P_\text{Total}}, \\
\label{eq:NLTPb}
\quad T(P) = \frac{P_{\text{Intra}}}{P_\text{Total}}.
\end{align}
\end{subequations}
\noindent

At this point it is important to remind that we use slow detectors for the power measurements. This means that the power is time-averaged over the pulse train. In particular, if the laser is in the period-two regime, average values of the low and high pulse energies are measured and one may ask if the details of the dynamic regime may be lost. As we will explain, the answer to this question is no.

We observe a curious phenomenon when the average powers are measured as function of the pump current: neither $P_{\text{Intra}}$ nor $P_{\text{Extra}}$ grows linearly with $I_{\text{pump}}$. This is in contradiction to what would be expected for a usual laser below saturation, for which the output power $P_{\text{Extra}}$ varies linearly with pump power. Instead, each curve presents a distinct behavior, especially in the period-four regime. In Fig.~\ref{f5}(a), for $I_{\text{pump}} < 829$ mA the extracavity and intracavity powers present opposite concavities. When $I_{\text{pump}}$ is increased above 829 mA, $P_{\text{Intra}}$ remains approximately constant while $P_{\text{Extra}}$ increases at a higher rate than previously. 

A possible explanation is that there is an abrupt variation in the distribution of instantaneous powers observed in the extracavity time series where the amplitude of some pulses increases faster with $I_{\text{pump}}$. The pulses with higher powers access regions of the PBS transmission curve where the nonlinear effects are more prominent and a description taking into account contributions beyond the first order power dependence, predicted in Eq.~(\ref{eq:NLR}), is necessary~\cite{Newbury}. This alters the shape of the effective transmission, the magnitude of the NPR and the measured average powers. Hence, the existence of pulses with higher powers in the extracavity signal are associated with a more pronounced NPR. 

By adding $P_{\text{Intra}}$ and $P_{\text{Extra}}$, another interesting effect becomes evident: the sum shows an approximately linear growth with $I_{\text{pump}}$, as seen by the curve $P_{\text{Total}}$ in Fig.~\ref{f5}(a). To appreciate the meaning of this observation we remind that $P_{\text{Total}}$ is the average power immediately after the gain section of the laser, just before the PBS. If one of the pulses in the pulse train has a higher peak power, its transmission through the PBS is higher but then the gain it experiences is smaller, due to gain saturation. The nonlinear contributions to gain saturation and nonlinear losses seem to compensate each other. The small signal (unsaturated) gain increases linearly with $I_{\text{pump}}$, and the net result is that the power after the gain section follows the small signal gain. In fact, the usual mode-locked operation requires this sort of balance to establish a stable pulse train. The difference in the present situation is that it does not occur on each individual pulse, but on the average pulse power, even during the pulse period doubling sequence.

The transition from period-two to period-four is also followed by changes in the extracavity optical spectrum, with the emergence of sidebands at 1025 nm, as shown in Fig.~\ref{f5}(b). These sidebands persist up to $I_{\text{pump}}$ = 969 mA and disappear when $I_{\text{pump}}<$ 829 mA, suggesting that it is associated with period quadrupling, when pulses with greater intensities arise. However, for the bifurcation around $I_{\text{pump}}$ = 545 mA, no transition was observed either in the average power or in the optical spectrum. This happens because, in the transition from regular regime to period-two, a small modulation gradually appears in the pulse intensities, whereas, in the bifurcation to period-four, occurs an abrupt change with immediate appearance of large pulses, making it possible to access nonlinearities previously unreachable. Larger intensities are associated with larger self-phase modulation, which broadens the optical spectrum. 

A clear experimental signature of the transitions between different dynamical regimes is shown in Fig.~\ref{f6}(a), where the nonlinear transmission $T$ as a function of $P_{\text{Total}}$ is presented. Note that the inclination of the curve changes around $P_{\text{Total}}$ = 165 mW (or $I_{\text{pump}}$ = 545 mA), which coincides with the period doubling. As pointed out, this bifurcation was not manifested in individual power measurements or in the optical spectrum, so that the nonlinear transmission curve brings more information and appears as a more complete characterization for the study of nonlinear dynamics of the laser. The transition to period-four occurs at $P_{\text{Total}}\approx$ 247 mW, which corresponds to $I_{\text{pump}}$ = 829 mA, in agreement with the previous results in Fig.~\ref{f5}. These transitions are better visualized in the derivative of $T$ with respect to $P_{Total}$ shown in Fig.~\ref{f6}(b), where abrupt changes occur for each bifurcation, making clear the existence of three different windows that relate to (I) regular mode-locked, (II) period-two and (III) period-four regimes. These discontinuities are a consequence of the changes in the NPR which occur for the different dynamical regimes. 

Many numerical models have been proposed to describe the dynamics of MLFL. However, accurate quantitative comparison between experiments and theory is difficult to achieve, due to the complexity of the physical system and the large number of parameters affecting the resulting dynamics~\cite{Wang:14, Wang:16}. Any of those models could be used to show that dynamical systems undergoing bifurcation display sudden variations in the average value of the dynamical variables~\cite{Hugo}. The average power in lasers is just another example, as seen for instance in the average power of far infrared gas lasers with saturable absorber~\cite{Rios}. In this last reference has been shown that the average power emitted by the laser has a sudden change in its slope as the pump power is scanned through bifurcations leading to different dynamical regimes. We show a similar phenomenon in our system in Fig.~\ref{f6}. 

\section{Conclusions}

In conclusion, we have measured the internal nonlinear transmission curve of an Yb MLFL. We have performed measurements while the laser performs transitions between the standard mode-locking, to period-two and period-four regime in the pulse energy. It is shown that this curve presents a clear experimental signature of the dynamical regime transitions where the slope suddenly varies as the regime changes. We have shown that the evolution of the intracavity and extracavity average powers are different, due to the variation of the NPR as the pump power increases. However, $P_{\text{Total}}$ presents a linear behavior, demonstrating complementary character of $P_{\text{Intra}}$ and $P_{\text{Extra}}$, which is a direct consequence of the energy conservation. As a next step, we are interested in extending the use of the nonlinear transmission curve to study harmonic mode-locking and multi-pulsing regimes and preliminary measurements in this direction have already been performed.

 This work was supported by the Brazilian research agencies CNPq (441668/2014-3, INCT-IQ 465469/2014-0), CAPES (PRONEX 534/2018, 23038.003382/2018-39), and FACEPE (APQ - 1178 - 1.05/14).



\end{document}